\documentclass[letterpaper, 10 pt, conference]{ieeeconf}  

\IEEEoverridecommandlockouts                              

\usepackage{cite}
\usepackage{amsmath,amssymb,amsfonts}
\usepackage{algorithmic}
\usepackage{graphicx}
\usepackage{textcomp}
\usepackage{xcolor}
\usepackage{subfigure}

\title{\LARGE \bf
Reduced Dynamics and Control for an Autonomous Bicycle*
}

\author{Jiaming Xiong$^{\$,1}$, Bo Li$^{\$,2}$, Ruihan Yu$^{1}$, Daolin Ma$^{\dag,3}$, Wei Wang$^{\dag,2}$, Caishan Liu$^{1}$
\thanks{*This work was performed under the support of the National Natural Science Foundation of China (NSFC:11932001,11702002).}
\thanks{$^1$ College of Engineering, Peking University.}
\thanks{$^2$ School of Mechanical Engineering and Automation, Beihang University.}
\thanks{$^3$ Mechanical Engineering Department, Massachusetts Institute of Technology.}
\thanks{$^\$$ Authors with equal contribution.}
\thanks{$^\dag$ Corresponding Authors. daolinma@mit.edu, wangwei701@buaa.edu.cn}
}

\begin{document}

\maketitle
\thispagestyle{empty}
\pagestyle{empty}

\begin{abstract}
In this paper, we propose the reduced model for the full dynamics of a bicycle and analyze its nonlinear behavior under a proportional control law for steering.
Based on the Gibbs-Appell equations for the Whipple bicycle, we obtain a second-order nonlinear ordinary differential equation (ODE) that governs the bicycle's controlled motion.
Two types of equilibrium points for the governing equation are found, which correspond to the bicycle's uniform straight forward and circular motions, respectively. By applying the Hurwitz criterion to the linearized equation, we find that the steer coefficient must be negative, consistent with the human's intuition of turning toward a fall. Under this condition, a critical angular velocity of the rear wheel exists, above which the uniform straight forward motion is stable, and slightly below which a pair of symmetrical stable uniform circular motions will occur. These theoretical findings are verified by both numerical simulations and experiments performed on a powered autonomous bicycle.
\end{abstract}


\section{Introduction}
A bicycle is a typical single-track vehicle. Compared to double-track vehicles, it is more efficient and maneuverable, especially in off-road environments such as deserts, mountains and forests~\cite{yi2006trajectory}. As a typical nonholonomic system, the bicycle has a nontrivial dynamical property: it is statically unstable but is easily stabilized or even self-stabilized at moderate speed. Studies on bicycle dynamics and control started at the end of the 19th century with the seminal work by Carvallo~\cite{carvallo1901theorie} and Whipple~\cite{whipple1899stability}, who derived a pair of linearized equations for the bicycle independently. Since then this topic has drawn attentions from many researchers in different areas. A detailed review of works on this topic can be found in~\cite{astrom2005bicycle,Meijaard2007,schwab2013review,kooijman2013review}.

Historically, there are three main points on the physical explanation of bicycle's self-stability, that is, the centrifugal force effect~\cite{timoshenko1948advanced}, the gyroscopic effect~\cite{klein1898theorie} and the caster effect~\cite{grammel1950}. 
However, it has become common belief that the key to bicycle's self-stability is turning toward a fall (TTF)~\cite{Kooijman2011}. Inspired by this stability rule, we adopt a proportional control law for the steering of an autonomous bicycle in this paper, i.e., the steer angle is controlled such that it is proportional to the lean angle by a steer coefficient. Simultaneously, the angular velocity of the rear wheel is controlled as a constant, for we mainly study the bicycle's uniform motion.

\begin{figure}[tbp]
\centering
\includegraphics[angle=0,width=5.5cm]{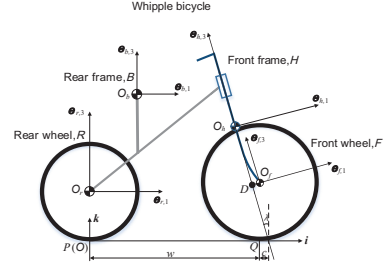}
\caption{\label{fig:1}
Whipple bicycle in an upright, straight reference configuration
}
\end{figure}

In the literature, the control of bicycle was always based on a simplified dynamics model, such as the inverted pendulum model~\cite{lowell1982}, the theoretical point mass model~\cite{Getz1995}, the double inverted pendulum model~\cite{keo2011experimental} and the linearized Carvallo-Whipple model~\cite{owczarkowski2016dynamic}. Using these models, one may control the bicycle to perform a stable straight forward motion, however, it is difficult to achieve the bicycle's circular motion or dynamical turning seen in biking races, due to the lack of the understanding of bicycle dynamics. Therefore, in order to improve the performance of the autonomous bicycle, a complete nonlinear dynamics model should be established. As presented in our previous work~\cite{xiong2019}, a good choice is to apply the Gibbs-Appell equations to the Whipple bicycle model, which is shown in Fig.~\ref{fig:1}. Compared to the above simplified models, this model is a reduced nonlinear multibody dynamics model where both the holonomic and nonholonomic constraints are embedded in.

Substituting the control law into this reduced dynamics model, we obtain a second-order ODE governing the controlled motion of the bicycle in this paper. We investigate the equilibrium points of the governing equation, which correspond to the bicycle's uniform straight forward motion (USFM) and uniform circular motion (UCM), respectively. Using the Hurwitz criterion, we find that the steer coefficient must be negative, agreeing with the stability rule of TTF. Under this condition, a supercritical pitchfork bifurcation occurs at a critical angular velocity of the rear wheel, above which the bicycle's USFM is stable. When the angular velocity is slightly less than the critical velocity, the USFM will lose stability, however, a pair of symmetrical stable UCMs will exist. The bicycle will truly loss stability when the angular velocity of the rear wheel is lower than another critical velocity, at which two saddle-node bifurcations occur. These theoretical findings are verified by both numerical simulations of the governing equation and experiments carried out on our powered autonomous bicycle.

The rest of this paper is organized as follows. We begin with a review of the related work in Sect.~\ref{related-work}. In Sect.~\ref{bicycle-dynamics}, we present a reduced dynamics model for the Whipple bicycle based on the Gibbs-Appell equations. The governing equation of the bicycle's controlled motion is derived in Sect.~\ref{control}, where both the trivial and nontrivial equilibrium points are studied. In Sect.~\ref{experiment}, we present the results of both numerical simulations and experiments performed on the powered autonomous bicycle. Conclusions are drawn in Sect.~\ref{conclusion}.

\section{Related work}\label{related-work}
In general, there are four main methods to stabilize and control the bicycle: steer control, control with a moving mass, control with a gyroscope, and a combination of these first three methods~\cite{schwab2013review,kooijman2013review}. Getz and Marsden~\cite{Getz1995} simplified the bicycle as a point mass with a zero tilt angle of the steering axis. They established the dynamics model using the Voronets equations, based on which they used steering and rear-wheel torque to control the bicycle to recover its balance from a near
fall as well as converge to a time parameterized path in the ground plane. In another paper, Getz~\cite{getz1995internal} applied an external/internal model decomposition approach to the problem of path-tracking with balance for the bicycle based on the same dynamics model. Tanaka and Murakami~\cite{Tanaka2004} derived an inverted pendulum model and proposed a PD control method for bicycle steering based on acceleration control. In order to realize the trajectory control for the bicycle, they proposed a lateral velocity controller and a steering function controller in their later work~\cite{Tanaka2009}. Keo and Yamakita~\cite{Keo2008,Keo2009} extended Getz's dynamics model by considering a nonzero tilt angle of the steering axis and adding a balancer to the bicycle's rear frame. They applied an output-zeroing controller and an input-output linearization approach to balance the bicycle and realize the trajectory tracking control, respectively. Keo et al.~\cite{keo2011experimental} performed experiments to verify a new balancer configuration for stabilizing of an unmanned bicycle based on output-zeroing controller. This balancer can be configured as a flywheel mode or a balancer mode by shifting the center of gravity of
the balancer. Yetkin et al.~\cite{Yetkin2014} designed a sliding mode controller to stabilize an unmanned bicycle with a control moment gyroscope at zero-forward velocity. Their dynamics model was derived using the Euler-Lagrange equation where both the kinetic and potential energies were simplified. Lot and Fleming~\cite{Lot2019} explored the potential of a gyroscopic stabilizer for the stabilization of single-track vehicles and found that the most effective configuration is one where the gyroscope spins with respect to an axis parallel to the wheel's spin axis and swings with respect to the vehicle yaw axis.

As mentioned above, most studies used simplified dynamics models of bicycle to implement control analysis, simulations and experiments. Chen et al.~\cite{Chen2006,Chen2010} claimed that these models were incapable of presenting all of the dynamic motions of the system in certain situations due to their simplicity. For this reason, they modelled the bicycle as a system with 11 generalized coordinates and 3 degrees of freedom in the velocity space using Lagrangian equations for quasi-coordinates, and then developed PID and fuzzy controllers for roll-angle-tracking. Similar works were done by Dao and Chen~\cite{Dao2011}, Chu and Chen~\cite{Chu2018} to use respectively sliding-mode control and model predictive control for roll-angle tracking.

The set of Lagrangian equations presented by Chen et al.~\cite{Chen2006,Chen2010} were completed, however, it introduced~8 Lagrangian multipliers, leading to~19 variables in total. The appearance of redundant variables will complicate the control analysis in terms of finding the equilibrium points and determining their stability. Spurious eigenvalues will occur in the stability analysis~\cite{xiong2019b}. With that in mind, we employ the Gibbs-Appell equations to obtain a reduced set of motion equations of the bicycle that are free of Lagrangian multipliers in this paper. Using these equations, the above problems can be well solved.

\section{Reduced dynamics of bicycle}\label{bicycle-dynamics}
In this section, we present a reduced dynamics model for the Whipple bicycle based on the Gibbs-Appell equations.

\subsection{Whipple bicycle model}
In this paper, we describe the system with the classical Whipple bicycle model~\cite{Meijaard2007}. As shown in Fig.~\ref{fig:1}, it consists of four rigid bodies: a rear wheel~$R$, a rear frame~$B$, a front frame~$H$ and a front wheel~$F$. Among which, the two wheels are circular symmetric and make ideal knife-edge rolling point contact with the horizontal ground, and the two frames have lateral symmetries in their shape and mass distributions. The effects of structural compliance, joint friction and rolling friction are assumed to be negligible. We define the following coordinate systems for the bicycle: an inertial frame~$\mathcal{F}_i=\{O;\boldsymbol{i},\boldsymbol{j},\boldsymbol{k}\}$, fixed
on ground, in which the~$\boldsymbol{i}$ and~$\boldsymbol{k}$ axes lie; and four body-fixed frames~$\mathcal{F}_r=\{O;\boldsymbol{e}_{r,1},\boldsymbol{e}_{r,2},\boldsymbol{e}_{r,3}\}$, $\mathcal{F}_b=\{O;\boldsymbol{e}_{b,1},\boldsymbol{e}_{b,2},\boldsymbol{e}_{b,3}\}$, $\mathcal{F}_h=\{O;\boldsymbol{e}_{h,1},\boldsymbol{e}_{h,2},\boldsymbol{e}_{h,3}\}$,
$\mathcal{F}_f=\{O;\boldsymbol{e}_{f,1},\boldsymbol{e}_{f,2},\boldsymbol{e}_{f,3}\}$, attached to the center of mass of each
body. The orientations of these frames are shown in Fig.~\ref{fig:1}.

Table~\ref{tab:1} lists the values of the parameters of our powered autonomous bicycle. There are~25 parameters, including the wheel base~$w$, the trail~$c$, the tilt angle~$\lambda$, the positions of the center of mass of the two frames~$x_k,z_k\;(k=b,h)$, the radii of the two wheels~$R_k\;(k=r,f)$, the masses of the four rigid bodies~$m_k\;(k=r,b,h,f)$, and the nonzero components of the inertia tensors of the four rigid bodies in their body-fixed frames~$I_{k,xx},I_{k,yy}\;(k=r,b,h,f)$ and~$I_{k,zz},I_{k,xz}\;(k=b,h)$.

\begin{table*}[htbp]
\centering
\caption{Values of the parameters of the powered autonomous bicycle:~$w=0.935\mathrm{m}$, $c=0.046\mathrm{m}$ and~$\lambda=0.175\mathrm{rad}$.}\label{tab:1}
\begin{tabular}{ccccccccc}
\hline \hline
Body \quad \quad&  $x_k$ & $z_k$ & $R_k$ & $m_k$ &  $I_{k,xx}$  & $I_{k,yy}$ & $I_{k,zz}$ & $I_{k,xz}$ \\
Rear wheel ($k=r$) \quad \quad& /\ & /\ & 0.260 & 1.0865 & 0.0293 & 0.0584 & /\ &  /\ \\
Rear frame ($k=b$) \quad \quad& 0.424 & 0.402 & /\ & 13.2490 & 0.2513  & 0.5147 & 0.3320 &  0.1215 \\
Front frame ($k=h$) \quad \quad& 0.865 & 0.554 & /\ & 2.8315 & 0.0365  & 0.0445 & 0.0132 &  -0.0157\\
Front wheel ($k=f$) \quad \quad& /\ & /\ & 0.260 & 1.0865 & 0.0293 & 0.0584 & /\ &  /\ \\
 \hline
 \hline
\end{tabular}\\
\footnotemark[1]{\footnotesize{Units: m for length, kg for mass, and $\rm {kg} \cdot {\rm m}^2$ for moment of inertia.}}
\end{table*}

Without any constraints, the generalized coordinates of the bicycle are chosen as~$\boldsymbol{q}=(x,y,z,\psi,\theta,\varphi,\delta,\phi_r,\phi_f)$. Here,~$(x,y,z)$ are the coordinates of the position of reference point~$D$, an intersection point of the steering axis and the coordinate axis~$O_f\boldsymbol{e}_{f,1}$, in the inertial frame~$\mathcal{F}_i$. The three Euler angles of the rear frame, $(\psi,\theta,\varphi)$, are defined by the~$312$ rotation sequence, representing the yaw, lean and pitch, respectively. Finally,~$\delta$ is the steer angle, and~$\phi_r$ and~$\phi_f$ are the rotation angles of the rear and front wheels, respectively. We point out that~$\theta$ is positive when leaning to the right, and~$\delta$ is positive when steering to the left.

\subsection{Gibbs-Appell equations}
When the bicycle moves on a horizontal ground, there are two holonomic constraints which take the following forms~\cite{xiong2019},
\begin{equation}\label{eq:1}
z=z(\theta,\delta),\;\varphi=\varphi(\theta,\delta).
\end{equation}
In the condition that no slip of the two wheels occur, we obtain four nonholonomic constraints, from which the generalized velocities~$\dot{\boldsymbol{q}}$ can be solved as linear combinations of three independent quasi velocities~$\dot{\boldsymbol{\sigma}}=(\dot{\sigma}^1,\dot{\sigma}^2,\dot{\sigma}^3)=(\dot{\theta},\dot{\delta},\dot{\phi}_r)$:
\begin{equation}\label{eq:2}
\dot{\boldsymbol{q}}=\boldsymbol{H}({\boldsymbol{q}})\dot{\boldsymbol{\sigma}}.
\end{equation}

Suppose that the steer and rear wheel torques exerted on the bicycle are~$\tau_\delta,\tau_{\phi_r}$, respectively. Denote by~$(\tau_1,\tau_2,\tau_3)=(0,\tau_\delta,\tau_{\phi_r})$. We employ the Gibbs-Appell equations to obtain a reduced set of motion equations of the bicycle that are free of Lagrangian multipliers. According to~\cite{xiong2019}, these equations take the following forms,
\begin{equation}\label{eq:3}
m_{ij}\ddot{\sigma}^j+c_{ijk}\dot{\sigma}^j\dot{\sigma}^k=P_i+\tau_i,\;i,j,k=1,2,3,
\end{equation}
where~$m_{ij},c_{ijk},P_i,\;i,j,k=1,2,3$ are functions of~$\theta,\delta$, i.e., $m_{ij}=m_{ij}(\theta,\delta),\,c_{ijk}=c_{ijk}(\theta,\delta),\,P_i=P_i(\theta,\delta)$, with the following properties:
\begin{equation}\label{eq:3a}
\left\{
\begin{aligned}
  &c_{i33}(0,0)=0,\;P_i(0,0)=0,\;i=1,2,    \\
  &c_{333}(\theta,\delta)\equiv0,\;P_3(\theta,\delta)=0.
\end{aligned}
\right.
\end{equation}

\section{Control law and stability}\label{control}
Inspired by TTF, we adopt the following control law for the bicycle,
\begin{equation}\label{eq:4}
\left\{
\begin{aligned}
  &\delta=c_1\theta,    \\
  &\dot{\phi}_r=\omega_0,
\end{aligned}
\right.
\end{equation}
where~$c_1$ is the steer coefficient and~$\omega_0$ is the given angular velocity of the rear wheel. The first and second equations of~(\ref{eq:4}) can be regarded as holonomic and nonholonomic constraint equations, respectively, and the steer and rear wheel torques are two corresponding constraint forces. Therefore, we can choose~$\theta$ as the only independent coordinate. Substituting~(\ref{eq:4}) into the first equation of~(\ref{eq:3}), we obtain a second-order ODE governing the controlled motion of the bicycle:
\begin{equation}\label{eq:5}
M_1^{c_1}(\theta)\ddot{\theta}+h_1^{c_1,\omega_0}(\theta,\dot{\theta})=0,
\end{equation}
where~$M_1^{c_1}=m_{11}(\theta,c_1\theta)+c_1 m_{12}(\theta,c_1\theta)$, and~$h_1^{c_1,\omega_0}$, the summation of terms~$c_{1jk}\dot{\sigma}^j\dot{\sigma}^k$ and~$-P_1$, is a function of~$\theta,\dot{\theta}$. Similarly, substituting~(\ref{eq:4}) into the second and third equations of~(\ref{eq:3}), we obtain the expressions of two torques~$\tau_\delta,\tau_{\phi_r}$:
\begin{equation}\label{eq:6}
\left\{
\begin{aligned}
  &\tau_\delta=M_2^{c_1}(\theta)\ddot{\theta}+h_2^{c_1,\omega_0}(\theta,\dot{\theta}),    \\
  &\tau_{\phi_r}=M_3^{c_1}(\theta)\ddot{\theta}+h_3^{c_1,\omega_0}(\theta,\dot{\theta}),
\end{aligned}
\right.
\end{equation}
where~$M_2^{c_1},M_3^{c_1}$ and~$h_2^{c_1,\omega_0},h_3^{c_1,\omega_0}$ are defined in the same way as~$M_1^{c_1}$ and~$h_1^{c_1,\omega_0}$.

\subsection{Stability of USFM}
According to~(\ref{eq:3a}), we know that~$(\theta,\dot{\theta})=(0,0)$ is a trivial equilibrium point of~(\ref{eq:5}) that corresponds to the bicycle's USFM, and the corresponding torques~$\tau_\delta,\tau_{\phi_r}$ are zero. The linearized equation of~(\ref{eq:5}) around~$(0,0)$ takes the following form:
\begin{equation}\label{eq:7}
a_2\ddot{\theta}+a_1\dot{\theta}+a_0\theta=0,
\end{equation}
where
\begin{equation}\label{eq:8}
\left\{
\begin{aligned}
  a_2=&\,m_{11}(0,0)+c_1m_{12}(0,0),   \\
  a_1=&\,c_1\omega_0\left(c_{123}(0,0)+c_{132}(0,0)\right),  \\
  a_0=&c_1c_{133,\delta}(0,0)\omega_0^2-P_{1,\theta}(0,0)-c_1P_{1,\delta}(0,0), \\
\end{aligned}
\right.
\end{equation}
and the subscripts~$\theta,\delta$ denote the derivatives with respect to them. In the derivation of~(\ref{eq:8}), we have used the fact that~$c_{113}(0,0)=c_{131}(0,0)=0$ and~$c_{133,\theta}(0,0)=0$, which can be verified via symbolic calculations. In addition, we list the expressions of~$c_{133,\delta}(0,0),P_{1,\theta}(0,0),P_{1,\delta}(0,0)$ as follows:
\begin{align}\label{eq:8a}
  c_{133,\delta}(0,0)=&-\frac{R_r\cos\lambda}{R_fw}\left(I_{f,yy}R_r + I_{r,yy}R_f + m_fR_rR_f^2 \right. \nonumber \\
  &\left.+m_rR_r^2R_f + m_bz_BR_rR_f + m_hz_HR_rR_f\right),     \nonumber \\
  P_{1,\theta}(0,0)=&\,(m_rR_r+m_fR_f+m_bz_b+m_hz_h)g,  \nonumber \\
  P_{1,\delta}(0,0)=&-\frac{gc\cos\lambda}{w}\left(m_bx_b+m_hx_h\right)-m_fgR_f\sin\lambda   \nonumber \\
  &+m_hg\left((w+c-x_h)\cos\lambda-z_h\sin\lambda\right),
\end{align}
where~$g=9.81\mbox{m}/\mbox{s}^2$ is gravity acceleration.

\begin{figure}[tbp]
\centering
\includegraphics[angle=0,width=6cm]{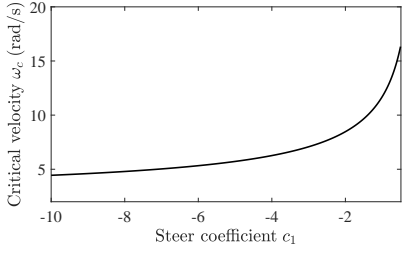}
\caption{\label{fig:2}
Curve of the critical velocity~$\omega_c$ with respect to the steer coefficient~$c_1$
}
\end{figure}

The equilibrium point~$(0,0)$ of~(\ref{eq:5}) is exponentially stable if and only if~(\ref{eq:7}) satisfies the Hurwitz criterion, that is, $a_0,a_1,a_2$ have the same sign. For the bicycle's parameters listed in Table~\ref{tab:1}, we find that~$a_2$ will be negative if and only if~$c_1$ is large enough, and it is not appropriate. Therefore, the two constants~$c_1,\omega_0$ should be chosen such that~$a_0,a_1,a_2$ are positive. Since~$c_{123}(0,0)=c_{132}(0,0)<0$ and~$\omega_0>0$, we have~$c_1<0$ due to the condition~$a_1>0$, agreeing
with the bicycle's stability rule of TTF. In this case, the condition~$a_2>0$ is always satisfied. In addition, since~$c_{133,\delta}(0,0)<0,\,P_{1,\theta}(0,0)>0,\,P_{1,\delta}(0,0)<0$, we obtain the range of~$\omega_0$ from the condition~$a_0>0$:
\begin{equation}\label{eq:9}
\omega_0>\sqrt{\frac{P_{1,\theta}(0,0)+c_1P_{1,\delta}(0,0)}{c_1c_{133,\delta}(0,0)}}:=\omega_c,
\end{equation}
where~$\omega_c$ is the critical velocity. According to~(\ref{eq:8a}), we know that~$\omega_c$ is independent of~$I_{k,xx}\;(k=r,f)$ and $I_{k,xx},I_{k,yy},I_{k,zz},I_{k,xz}\;(k=b,h)$. Using the bicycle's parameters listed in Table~\ref{tab:1}, we plot the curve of~$\omega_c$ with respect to~$c_1$ in the range of~$c_1\in[-10,-0.5]$ in Fig.~\ref{fig:2}. The figure shows that the critical velocity of the bicycle increases with the growing of~$c_1$, which means, with milder steer coefficient, the bicycle's speed needs to increase in order to run stably in a straight line. As~$c_1$ tends to~$-\infty$, we know from~(\ref{eq:9}) that~$\omega_c$ tends to~$\sqrt{P_{1,\delta}(0,0)/c_{133,\delta}(0,0)}$, the lowest velocity for the bicycle's USFM under the control law~(\ref{eq:4}).

\subsection{Global analysis}\label{Globalanalysis}
The stability of~$(0,0)$ discussed in the above section is in a local sense. To estimate its basin of attraction, one may use the Lyapunov's direct method for a global analysis. However, it is difficult to construct a Lyapunov function due to the high nonlinearity of~(\ref{eq:5}).
Further investigation of~(\ref{eq:5}) shows that nontrivial equilibrium points may exist under certain values of~$c_1$ and~$\omega_0$, which correspond to the bicycle's UCMs. We point out that for a stable~$(0,0)$, the distance between it and a nearby unstable nontrivial equilibrium point
can give an estimation for the size of its basin of attraction.

Denote by~$(\theta_0,0)$ an equilibrium point of~(\ref{eq:5}), where~$\theta_0$ satisfies~$h_1^{c_1,\omega_0}(\theta_0,0)=0$, and thus is determined by the following equation:
\begin{equation}\label{eq:10}
c_{133}(\theta_0,c_1\theta_0)\omega_0^2-P_1(\theta_0,c_1\theta_0)=0.
\end{equation}
Due to the lateral symmetry of the bicycle, if~$(\theta_0,0)$ is an equilibrium point, so is~$(-\theta_0,0)$. The linearized equation around~$(\theta_0,0)$ takes the following form:
\begin{equation}\label{eq:11}
b_2\Delta\ddot{\theta}+b_1\Delta\dot{\theta}+b_0\Delta\theta=0,
\end{equation}
where~$\Delta\theta=\theta-\theta_0$ and
\begin{equation}\label{eq:12}
\left\{
\begin{aligned}
  b_2=&\,m_{11}(\theta_0,c_1\theta_0)+c_1m_{12}(\theta_0,c_1\theta_0),   \\
  b_1=&\,c_1\omega_0\left(c_{123}(\theta_0,c_1\theta_0)+c_{132}(\theta_0,c_1\theta_0)\right)   \\
  &+\left(c_{113}(\theta_0,c_1\theta_0)+c_{131}(\theta_0,c_1\theta_0)\right)\omega_0,   \\
  b_0=&\left(c_{133,\theta}(\theta_0,c_1\theta_0)+c_1c_{133,\delta}(\theta_0,c_1\theta_0)\right)\omega_0^2 \\ &-P_{1,\theta}(\theta_0,c_1\theta_0)-c_1P_{1,\delta}(\theta_0,c_1\theta_0). \\
\end{aligned}
\right.
\end{equation}
According to the Hurwitz criterion,~$(\theta_0,0)$ is exponentially stable if and only if~$b_0,b_1,b_2$ have the same sign.

\begin{figure}[tbp]
\centering
\includegraphics[angle=0,width=6cm]{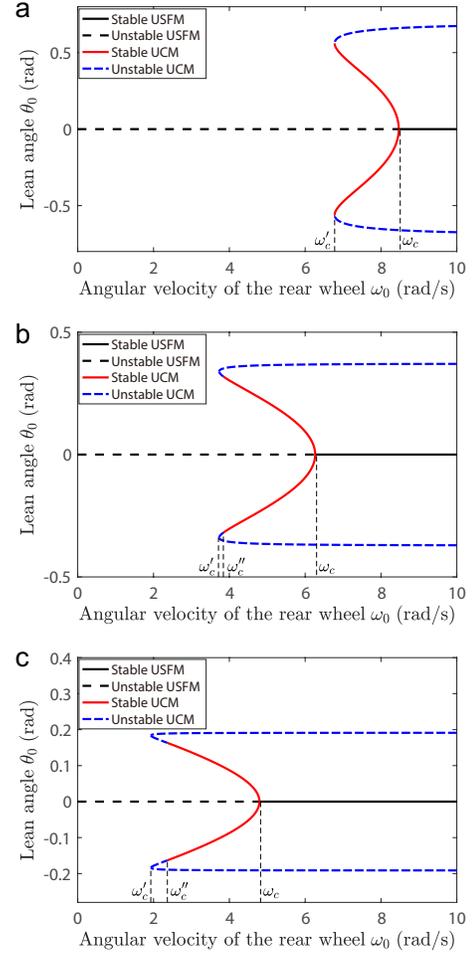}
\caption{\label{fig:3}
Bifurcation diagrams
}
\end{figure}

As shown in Fig.~\ref{fig:3}a-\ref{fig:3}c, by fixing~$c_1=-2,-4,-8$, respectively, we obtain three bifurcation diagrams where~$\omega_0$ is the only parameter. In general, these diagrams show the same pattern, that is, a supercritical pitchfork bifurcation occurs at~$\omega_0=\omega_c$ and two saddle-node bifurcations occur at~$\omega_0=\omega_c^\prime$.
Unlike~$\omega_c$, the critical velocity~$\omega_c^\prime$ does not have analytical expression. However, it can be numerically determined.
The values of~$\omega_c$ and~$\omega_c^\prime$ decrease as the absolute value of~$c_1$ increases. In each bifurcation diagram, points on the black solid line, black dotted line, red solid line and blue dotted line represent the bicycle's stable USFM, unstable USFM, stable UCM and unstable UCM, respectively.

In the case of~$c_1=-2$, when~$\omega_0>\omega_c$, (\ref{eq:5}) has three equilibrium points: the stable equilibrium point~$(0,0)$ and a pair of symmetrical unstable nontrivial equilibrium points.
When~$\omega_c^\prime<\omega_0<\omega_c$, (\ref{eq:5}) has five equilibrium points: the unstable equilibrium point~$(0,0)$, a pair of symmetrical unstable nontrivial equilibrium points and a pair of symmetrical stable nontrivial equilibrium points. In this case, the bicycle's USFM will lose stability, however, a pair of symmetrical stable UCMs will exist. This is an interesting phenomenon. When~$\omega_0<\omega_c^\prime$, (\ref{eq:5}) has only one equilibrium point: the unstable equilibrium point~$(0,0)$. So in this case, the bicycle will truly lose stability.

In the case of~$c_1=-4$, there is a little difference in the stability of equilibrium points. We can see that another critical velocity~$\omega_{c}^{\prime\prime}\in(\omega_{c}^{\prime},\omega_c)$ exists.
When~$\omega_{c}^{\prime\prime}<\omega_0<\omega_c$, (\ref{eq:5}) still has a pair of stable equilibrium points. However, when~$\omega_{c}^{\prime}<\omega_0<\omega_c^{\prime\prime}$, these equilibrium points lose stability. This phenomenon is also observed in the case of~$c_1=-8$. We find that as the absolute value of~$c_1$ increases, the distance between~$\omega_{c}^\prime$ and~$\omega_{c}^{\prime\prime}$ becomes larger.

\section{Experiment}\label{experiment}
In order to verify the theoretical findings, we carry out both numerical and real experiments in this section.

\subsection{Numerical experiment}\label{simulation}
\begin{figure}[tbp]
\centering
\includegraphics[angle=0,width=6cm]{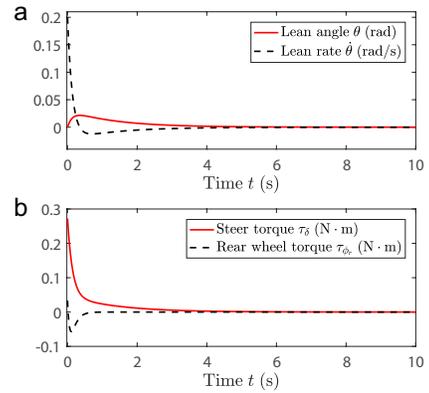}
\caption{\label{fig:4}
First numerical example, $c_1=-4,\,\omega_0=7\mathrm{rad}/\mathrm{s}$
}
\end{figure}

First, we present two numerical examples with the same steer coefficient~$c_1=-4$ and initial condition $(\theta(0),\dot{\theta}(0))=(0\mathrm{rad},0.2\mathrm{rad}/\mathrm{s})$. The corresponding bifurcation points are~$\omega_c=6.26\mathrm{rad}/\mathrm{s}$, $\omega_c^\prime=3.73\mathrm{rad}/\mathrm{s}$ and~$\omega_c^{\prime\prime}=3.84\mathrm{rad}/\mathrm{s}$, respectively. In the first example, we choose~$\omega_0=7\mathrm{rad}/\mathrm{s}>\omega_c$. In this case, (\ref{eq:5}) has an exponentially stable equilibrium point~$(0,0)$. The numerical solution of~(\ref{eq:5}) is shown in Fig.~\ref{fig:4}a. We see that both~$\theta$ and~$\dot{\theta}$ will quickly converge to~$0$. Fig.~\ref{fig:4}b shows the evolutions of the steer and rear wheel torques with respect to time. As~$t$ tends to~$\infty$, both torques will tend to~$0$.

\begin{figure}[tbp]
\centering
\includegraphics[angle=0,width=6cm]{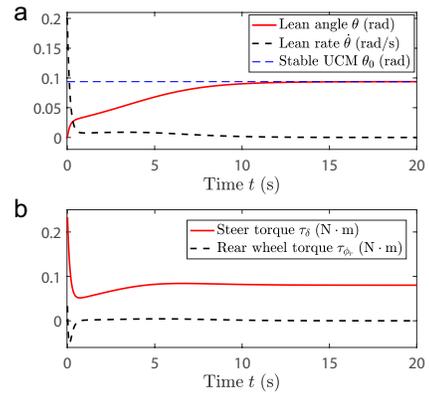}
\caption{\label{fig:5}
Second numerical example, $c_1=-4,\,\omega_0=6\mathrm{rad}/\mathrm{s}$
}
\end{figure}

For the second example, we choose~$\omega_0=6\mathrm{rad}/\mathrm{s}\in(\omega_c^{\prime\prime},\omega_c)$. According to Fig.~\ref{fig:3}b, $(0,0)$ is unstable in this case, but two stable nontrivial equilibrium points~$(\pm\theta_0,0)$ exist, where~$\theta_0=0.0938\mathrm{rad}$. The numerical solution of~(\ref{eq:5}) is shown in Fig.~\ref{fig:5}a.
Clearly, it converges to the equilibrium point~$(\theta_0,0)$, which is represented by the blue dotted line.
Fig.~\ref{fig:5}b shows the evolutions of the steer and rear wheel torques with respect to time. As~$t$ tends to~$\infty$, $\tau_{\phi_r}$ will tend to zero, but $\tau_{\delta}$ will tend to a nonzero value.

\subsection{Real Experiment}
\begin{figure}[tbp]
\centering
\includegraphics[angle=0,width=7.cm]{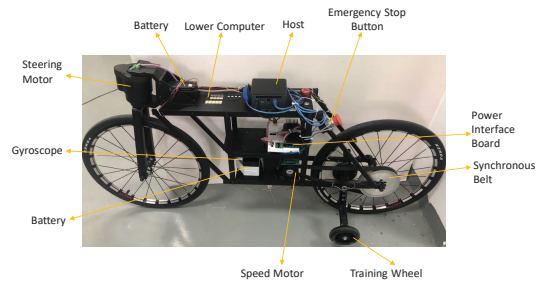}
\caption{\label{fig:8}
Bicycle hardware system
}
\end{figure}

We carry out real experiments using our powered autonomous bicycle, which is shown in Fig.~\ref{fig:8}. On the basis of the bicycle's rear frame, we add a single-axis fiber optical gyroscope to measure the lean angle and lean rate. In order to process the angle information and run the control algorithm, we need a small PC. We use two servo motors to control the speed of the rear wheel and the steering of the handlebars, respectively. The speed motor adopts the built-in motor of the electric vehicle sold on the market, while its motor drive is replaced with a 7020Su drive. The steering motor uses the same drive, and its power is~$50\mbox{W}$, with a maximum angular velocity of~$60^\circ/\mbox{s}$. We use the maximum speed to control the rotation of the handlebar in the experiment. In order to establish good communication for fast and effective control, we use a high-performance controller with stm32-bit core. It can receive the control commands sent by the PC, and control the movement of the two motors separately with the RS485 communication protocol.

\begin{figure}[tbp]
\centering
\includegraphics[angle=0,width=6cm]{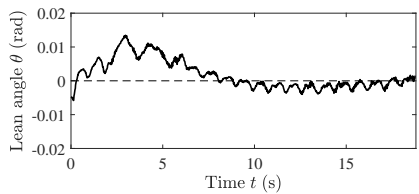}
\caption{\label{fig:6}
Evolution of the lean angle in the case of $c_1=-4$ and~$\omega_0=7\mathrm{rad}/\mathrm{s}$
}
\end{figure}

\begin{figure}[tbp]
\centering
\includegraphics[angle=0,width=5cm]{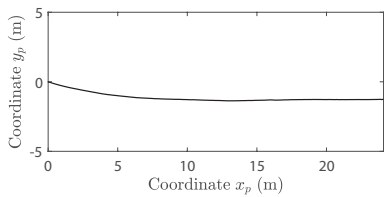}
\caption{\label{fig:9}
Trajectory of the marker point on the $x-y$ plane in the case of $c_1=-4$ and~$\omega_0=7\mathrm{rad}/\mathrm{s}$
}
\end{figure}

First, we verify the bicycle's stable USFM under the condition that its velocity is greater than the critical velocity~$\omega_c$. Fig.~\ref{fig:6} shows the evolution of the lean angle with respect to time in the case of~$c_1=-4$ and~$\omega_0=7\mathrm{rad}/\mathrm{s}$. Clearly, the evolution has the similar tendency as that in the first numerical example in Sect.~\ref{simulation} (see Fig.~\ref{fig:4}a). We can see that after an obvious perturbation in the first eight seconds, the lean angle converges to~$0$ with small oscillation, meaning that the bicycle moves forward near the upright straight configuration. In addition, we select a marker point on the bicycle's rear frame. The evolutions of its coordinates~$(x_p(t),y_p(t))$ on the~$x-y$ plane can be identified from the images taken by the unmanned aerial vehicle, as shown in Fig.~\ref{fig:9}, where the initial coordinates are~$(x_p(0),y_p(0))=(0,0)$. We can see that the trajectory converges to a straight line.
In summary, the result of the experiment validates the theoretical prediction.

\begin{figure}[tbp]
\centering
\includegraphics[angle=0,width=6cm]{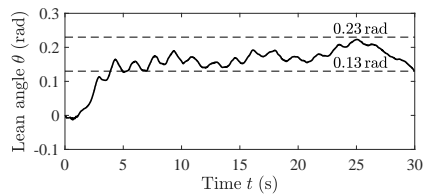}
\caption{\label{fig:7}
Evolution of the lean angle in the case of $c_1=-4$ and~$\omega_0=6\mathrm{rad}/\mathrm{s}$
}
\end{figure}

\begin{figure}[tbp]
\centering
\includegraphics[angle=0,width=4.3cm]{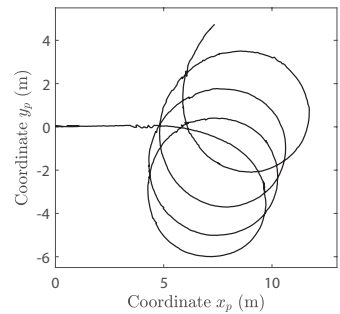}
\caption{\label{fig:10}
Trajectory of the marker point on the $x-y$ plane in the case of $c_1=-4$ and~$\omega_0=6\mathrm{rad}/\mathrm{s}$
}
\end{figure}

Then, we verify the bicycle's stable UCM under the condition that its velocity is slightly less than the critical velocity~$\omega_c$. Fig.~\ref{fig:7} shows the evolution of the lean angle with respect to time in the case of~$c_1=-4$ and~$\omega_0=6\mathrm{rad}/\mathrm{s}$. Generally, the evolution of lean angle shows the similar pattern as that in the second numerical example in Sect.~\ref{simulation} (see Fig.~\ref{fig:5}a). The initial value of the lean angle is close to~$0$. But since~$(0,0)$ is an unstable equilibrium point of~(\ref{eq:5}) in this case, the lean angle will deviate from~$0$. It starts to oscillate within the range of~$(0.13\mbox{rad},0.23\mbox{rad})$ as~$t>5\mbox{s}$, meaning that the bicycle moves forward in a persistent turning motion to the same side. However, since the lean angle does not tend to a constant value, this motion is not an exact UCM. As shown in Fig.~\ref{fig:10}, the trajectory of the marker point on the~$x-y$ plane is not a perfect closed circle.
The difference between the experimental and theoretical results is believed to come from the measurement error of the gyroscope. We find that this error accumulates as the yaw angle of the bicycle increases. To obtain a quantitative analysis for the influence of the error on the bicycle's controlled motion, an augmented model with a constant term included in the first equation of the control law~(\ref{eq:4}) should be developed in the future.
In general, despite these differences, the result of this experiment also confirms the validity of the theoretical model.

\section{Conclusion and discussion}\label{conclusion}
This paper presents a reduced model for full dynamics of the Whipple bicycle. A proportional control law is proposed to regulate the bicycle's motion with the inspiration of turning toward a fall. We analyze the nonlinear dynamics of the bicycle under this control law. The equilibrium points correspond to two typical motions, namely USFM and UCM, and their exponentially stability can be studied using the Hurwitz criterion.
Under the condition that the steer coefficient is negative, we find that a critical angular velocity of the rear wheel exists, above which the USFM is stable, and slightly below which a pair of symmetrical stable UCMs will occur. Further slowing the bicycle's velocity will result in the true unstability that indicates abrupt falls.

For real experiment, we design a powered autonomous bicycle equipped with a gyroscope and actuation motors at both the steering joint and rear wheel. The above theoretical findings are verified by both numerical simulations and real experiments.
In the future, we will analyze the robustness of the proposed control law in the presence of impulsive external disturbances.

In general, this paper reflects the importance of a complete dynamics model for the implementation of control analysis, simulations and experiments. The reduced dynamics model presented in this paper enables the analysis of bicycle's complete dynamics and is promising for designing more complicated and effective control algorithms to improve the performance of the autonomous bicycle in more challenging scenarios such as stable ultra-low-speed USFM, efficient dynamical turning and complex trajectory tracking.


\bibliographystyle{IEEEtran}
\bibliography{bicycledynamics}

\vspace{12pt}

\end{document}